\documentclass[pre,aps,floatfix,twocolumn,superscriptaddress,showpacs]{revtex4-1}
\usepackage{color}
\usepackage{graphicx}
\usepackage{longtable}
\usepackage{amsmath}
\usepackage{amsthm}
\usepackage{amssymb}
\usepackage{lipsum}
\usepackage{latexsym}
\usepackage{bm}
\newcommand{\mbf}{\mathbf}

\definecolor{lightblue}{rgb}{.0, .0, .8}

\graphicspath{{./figures/}}

\begin{document}
\date{\today}
\title{Spatial correlations of hydrodynamic fluctuations in simple fluids under shear flow: A mesoscale simulation study}
\author{Anoop Varghese}
\affiliation{Institute of Complex Systems and Institute for Advanced Simulation, Forschungszentrum J\"{u}lich, J\"{u}lich 52425, Germany}
\affiliation{School of Mathematical, Physical, and Computational Sciences, University of Reading, Whiteknights, Reading RG6 6AX, United Kingdom}
\author{Gerhard Gompper}
\affiliation{Institute of Complex Systems and Institute for Advanced Simulation, Forschungszentrum J\"{u}lich, J\"{u}lich 52425, Germany}
\author{Roland G. Winkler}
\affiliation{Institute of Complex Systems and Institute for Advanced Simulation, Forschungszentrum J\"{u}lich, J\"{u}lich 52425, Germany}
\begin{abstract}
Hydrodynamic fluctuations in simple fluids under shear flow are demonstrated to be spatially correlated, in contrast to the fluctuations at equilibrium, using
mesoscopic hydrodynamic simulations.
The simulation results for the equal-time hydrodynamic correlations in a multiparticle collision dynamics (MPC) fluid in shear flow are compared with the
explicit expressions obtained from fluctuating hydrodynamic calculations.
For large wave vectors $k$, the nonequlibrium contributions to transverse and longitudinal velocity
correlations decay as $k^{-4}$ for wave vectors along the flow direction, and as $k^{-2}$ for the off-flow directions.
For small wave vectors, a cross-over to a slower decay occurs,  indicating long-range correlations in real space.
The coupling between the transverse velocity components, which vanishes
at equilibrium, also exhibits a $k^{-2}$ dependence on the wave vector.
In addition, we observe a quadratic dependency on the shear rate of the non-equilibrium contribution to pressure.
\end{abstract}
\maketitle

\section{Introduction}

Fluctuations are an integral part of static and dynamic properties of a fluid.
At equilibrium, characteristic properties such as viscosity, compressibility, thermal diffusivity etc.
are often calculated from correlations in the hydrodynamic fluctuations~\cite{boon_book}.
The hydrodynamic fluctuations in simple fluids at equilibrium are well known to be spatially short-ranged \cite{hansen_book}.
However, in non-equilibrium steady-states they are in general long-ranged~\cite{zarate_book}.
This long-range correlations are present even in the absence of phase
transitions and hydrodynamic instabilities. The long-range nature of hydrodynamic fluctuations in nonequilibrium steady states was first observed
by mode-coupling theories as the enhancement of the Rayleigh component of the structure factor under temperature gradients~\cite{kirkpatrick1980}.
This observation was complemented by fluctuating hydrodynamic calculations~\cite{ronis1982}, and was also confirmed
in light scattering experiments~\cite{law1988,segre1992}.

Hydrodynamic fluctuations in shear flow, similar to those in temperature gradients,
are distinctively different from the fluctuations at equilibrium~\cite{lutsko1985,wada2003,zarate2008,zarate2009,otsuki2009-pre,otsuki2009-epj}.
In Ref.~\cite{lutsko1985}, using fluctuating hydrodynamics calculations,  the correlations in hydrodynamic fluctuations in a sheared fluid were evaluated
for small wave vectors and arbitrary shear rates. It was shown that the temporal decay of the velocity and density fluctuations in shear flow are anisotropic and shear-rate dependent, which we recently
verified using mesoscopic hydrodynamic simulations~\cite{varghese2015}.
More importantly, the equal-time correlations were predicted to be spatially long-ranged and follow an algebraic decay for unbounded systems.
The algebraic decay of the correlations in shear flow shares some similarities with that in temperature gradients, albeit the
origin of the long-range character is apparently different~\cite{zarate2004}.
The long-range nature of the correlations may also manifest as non-intensivity of the nonequilibrium contribution
to pressure~\cite{wada2003}.
Moreover, the spatial  correlations imply that the long-wavelength components of the fluctuations are strongly affected by the presence of confining walls~\cite{zarate2008,zarate2009,zarate2011,zarate2013}.

The nonequilibrium contribution to the hydrodynamic fluctuations in shear flow under normal experimental conditions is much weaker than that in
temperature gradients~\cite{lutsko1985,wada2003,zarate2008}. As a result, very little is known about
the long-range nature of hydrodynamic correlations from shear flow experiments.
However, computer simulations provide  alternative testing grounds for hydrodynamic theories.
In this paper, we show by multiparticle collision dynamics (MPC) simulations, a mesoscopic hydrodynamic simulation method \cite{kapral2008,gompper2009}, that  hydrodynamic fluctuations in shear flow are indeed
spatially correlated. In particular, the wave-vector dependence of the nonequilibrium contribution to velocity and density fluctuations are elucidated.
To this end, we derive the explicit analytical expressions for the equal-time correlation functions in
an isothermal MPC fluid under shear flow, and confirm the theoretical predictions by simulations.
Our calculations are based on that in Ref.~\cite{lutsko1985}, extending them to isothermal fluids with an asymmetric stress tensor.

This paper is organized as follows. In Sec.~\ref{hi}, the linearized Landau-Lifshitz Navier-Stokes equations for an MPC fluid under shear flow are derived.
In Sec.~\ref{correlations}, the equal-time hydrodynamic correlation functions are defined and explicit expressions for the non-equilibrium contributions are obtained.
The details of the MPC simulations are given in Sec.~\ref{simulations}. The comparison between theoretical and simulation results for the velocity and density correlations are given in Secs.~\ref{simulation_velocity}
and \ref{simulation_density}. The simulation results for the non-equilibrium contribution to the pressure are given in Sec.~\ref{simulation_pressure}. Conclusions and discussions are given in Sec.~\ref{conclusions}

\section{Theory}
\subsection{Linearized Landau-Lifshitz Navier-Stokes equation of MPC fluid}{\label{hi}}
We consider the non-angular-momentum conserving variant of a MPC fluid, which is characterized by the asymmetric stress tensor~\cite{pooley2005,gompper2009,winkler2009}
\begin{equation}
\sigma_{\alpha\beta}
=\eta^k\left[\frac{\partial u_\alpha}{\partial r_\beta}+\frac{\partial u_\beta}{\partial r_\alpha}-\frac{2}{3}\delta_{\alpha\beta}\frac{\partial u_\delta}{\partial r_\delta}\right]
+\eta^c\frac{\partial u_\alpha}{\partial r_\beta}~,
\label{sigma_ij}
\end{equation}
where $\eta^k$ and $\eta^c$ are the kinetic and collisional part of the viscosity.
Here, the Greek indices denote the Cartesian coordinates, and the Einstein summation convention is applied.
We consider isothermal fluid, where the temperature fluctuations decay at a shorter time scale compared to density and velocity fluctuations, so that
the dynamics of the two sets are decoupled and the temperature can be taken as constant~\cite{hijar2011,huang2015}.
The evolution of the density $\rho$ and velocity $\mathbf u$ of the fluid are then given by the Landau-Lifshitz Navier-Stokes equations
\begin{align} \label{mass_eqn}
\frac{\partial \rho}{\partial t} & = -\nabla\cdot(\rho \bf u)~, \\
\rho \left[ \frac{\partial}{\partial t} + \mathbf u \cdot \nabla \right]\mathbf u & = -\nabla p + \eta \nabla^2
\mathbf u + \frac{\eta ^k}{3}\nabla \left (\nabla\cdot \mathbf u\right )+\mathbf f^R ,
\label{NS_eqn}
\end{align}
with the viscosity $\eta=\eta^k+\eta^c$, the random force $\mathbf f^R=\nabla \cdot \boldsymbol \sigma^R$, and the fluctuating part  $\boldsymbol\sigma^R$ of the stress tensor.
The fluctuations $\boldsymbol \sigma^R$ obey the fluctuation-dissipation relation~\cite{landau_book}
\begin{equation}
\langle \sigma^R_{\alpha\beta}(\mathbf r,t)\sigma^R_{\gamma\delta}(\mathbf r',t')\rangle=2k_BT\eta_{\alpha\beta\gamma\delta}\delta(\mathbf r-\mathbf r')\delta(t-t') .
\label{FDT}
\end{equation}
The viscosity coefficients $\eta_{\alpha\beta\gamma\delta}$ are related to the stress tensor through the constitutive relation
$\sigma_{\alpha\beta}=\eta_{\alpha\beta\gamma\delta}\frac{\partial u_\gamma}{\partial r_\delta}$~\cite{landau_book}.
Using the explicit form of $\sigma_{\alpha\beta}$ from Eq.~(\ref{sigma_ij}), we get
\begin{equation}
\eta_{\alpha\beta\gamma\delta}=\eta \delta_{\alpha\gamma}\delta_{\beta\delta}+\eta^k\delta_{\alpha\delta}\delta_{\beta\gamma}-\frac{2}{3}\eta^k\delta_{\alpha\beta}\delta_{\gamma\delta}~.
\label{eta_ijkl}
\end{equation}
Equations~(\ref{mass_eqn}) and (\ref{NS_eqn}) are linearized by setting $\rho=\rho_0+\delta \rho$, $p=p_0+\delta p$, and $\mathbf u=\mathbf u_0+\delta \mathbf u$, where the mean flow velocity is
$u_{0\alpha}=\dot \gamma_{\alpha\beta} r_\beta$ with the shear rate tensor $\dot \gamma_{\alpha\beta}$.
We choose $\hat{\mathbf x}$ as the flow direction and $\hat{\mathbf y}$ as the gradient direction (the  circumflex indicates unit vectors),
such that $\dot \gamma_{\alpha\beta}=\dot \gamma \delta_{\alpha x}\delta_{\beta y}$, where  $\dot \gamma$ is the shear rate. The
MPC fluid is characterized by the ideal-gas equation of state \cite{huang2015,kapral2008,gompper2009}, and therefore the pressure fluctuations at constant temperature are given by $\delta p=c_T^2 \delta \rho$, where 
$c_T$ is the isothermal speed of sound. On linearizing, equations~(\ref{mass_eqn}) and (\ref{NS_eqn}) can then be written in the Fourier space as
~\cite{lutsko1985,varghese2015}
\begin{equation}
 \frac{\partial \tilde{\mathbf z}}{\partial t}+\left[-\dot\gamma k_x\frac{\partial}{\partial k_y}+\mathcal L(\mathbf k,\dot\gamma) \right] \tilde{\mathbf z}=\tilde{\mathbf R}~,
\label{eq10}
\end{equation}
where $\tilde{\mathbf z}=(\delta \tilde{\rho},\delta \tilde u^{\left(1\right)},\delta \tilde u^{\left(2\right)}, \delta \tilde u^{\left(3\right)})$, with
$\tilde u^{\left(\alpha\right)}=\delta \tilde{\mathbf u}(\mathbf k)\cdot \mathbf e^{\left(\alpha\right)}(\mathbf k)$,
$\tilde R_{\alpha+1}=\tilde{\mathbf f}\cdot\mathbf e^{\left(\alpha\right)}$, and $\tilde R_1=0$. Here, $\{\mathbf e^{(\alpha)}\}$
is a set of orthogonal unit vectors with  $\mathbf e^{(1)}$ pointing along the wave vector.
Thereby, $\delta \tilde u^{(1)}$ is the longitudinal, and $\delta\tilde u^{(2)}$ and $\delta\tilde u^{(3)}$
are the transverse components of the velocity fluctuations. Here, functions with a tilde indicate the Fourier components of the hydrodynamic fields and random forces.
As in Ref.~\cite{lutsko1985}, we choose the unit vectors as
\begin{eqnarray}\nonumber
\mathbf e^{\left(1\right)}&=&\hat{\mathbf k} , \\
{\mathbf e}^{\left(2\right)}&=&\frac{1}{\hat{k}_\perp}\left[\hat{\mathbf y}-{\mathbf e}^{\left(1\right)}e^{\left(1\right)}_y\right] , \\
{\mathbf e}^{\left(3\right)}&=& {\mathbf e}^{\left(1\right)}\times {\mathbf e}^{\left(2\right)}\nonumber
\end{eqnarray}
for which the matrix $\mathcal L(\mathbf k,\dot\gamma)$ takes a simple form. Here, $\hat k_\perp=(k_x^2+k_z^2)^{1/2}/k$ and $k=|\mathbf k|$.
The explicit form of $\mathcal L(\mathbf k,\dot\gamma)$ for an isothermal MPC fluid is presented in Ref.~\cite{varghese2015}.
The solution of Eq.~(\ref{eq10}) can then be written as~\cite{varghese2015,otsuki2009-epj}
\begin{align}
\tilde z_{i}(\mathbf k,t)=&~\sum_{j=1}^{4}G_{ij}(\mathbf k,t)\tilde z_{j}(\mathbf k(-t),0)\nonumber \\
&+\sum_{j=1}^{4}\int_0^tdt' G_{ij}(\mathbf k,t')\tilde R_j(\mathbf k(-t'),t-t') ,
\label{solution}
\end{align}
where $i\in\{1,2,3,4\}$ and the propagator $G_{ij}(\mathbf k,t)$ is given by
\begin{equation}
G_{ij}(\mathbf k,t)=\sum_{l=1}^{4}\xi^{\left(l\right)}_i(\mathbf k)\eta^{\left(l\right)}_j(\mathbf k(-t))e^{-\int_0^t d\tau \lambda_l(\mathbf k(-\tau))}~.
\label{propogator}
\end{equation}
Here, the time-dependent wave vector is defined as $\mathbf k(t)=(k_x,k_y-\dot{\gamma}t k_x,k_z)$, and $\boldsymbol \xi^{\left(l\right)}$ and $\boldsymbol \eta^{\left(l\right)}$
are the right and left eigenvectors of the operator $-\dot\gamma k_x \partial/\partial k_y+\mathcal L$,
corresponding to eigenvalue $\lambda_l$. In addition, $\sum_{l=1}^{4}\xi^{\left(l\right)}_i\eta^{\left(l\right)}_j=\delta_{ij}$, so  that $\{\boldsymbol \xi,\boldsymbol \eta\}$
forms a biorthogonal basis. The eigenvectors and eigenvalues can be obtained perturbatively as expansions in the wave vector~\cite{lutsko1985}. The explicit expressions are
given in Ref.~\cite{varghese2015}.

\subsection{Correlation functions}{\label{correlations}}
The steady state equal-time correlation functions $C_{ij}(\mathbf k,\dot\gamma)$ of the hydrodynamic variables are defined by
\begin{equation}
\lim_{t\to\infty}\langle\tilde z_{i}(\mathbf k,t)\tilde z_j(\mathbf k',t) \rangle = \left(2\pi\right)^3\delta(\mathbf k+\mathbf k')C_{ij}(\mathbf k,\dot\gamma)~.
\end{equation}
From Eqs.~(\ref{solution}) and (\ref{propogator}), and using the condition $G_{ij}(\mathbf k,t)=0$ for $t\rightarrow\infty$, we obtain
\begin{equation}
C_{ij}(\mathbf k,\dot{\gamma})=\sum_{l,m=1}^{4}\int_0^\infty dt~\xi^{\left(l\right)}_i(\mathbf k,t)\xi^{(m)}_j(-\mathbf k,t)F^{\left(lm\right)}(\mathbf k(-t)) ,
\label{c_ij_def}
\end{equation}
where we have used the definitions
\begin{align}
\xi^{\left(l\right)}_i(\mathbf k,t)&=\xi^{\left(l\right)}_i(\mathbf k)e^{-\int_0^t d\tau \lambda_l(\mathbf k(-\tau))} ,\\
\langle \tilde R_{i}(\mathbf k,t)\tilde R_j(\mathbf k',t')\rangle&=(2\pi)^3\delta(\mathbf k+\mathbf k')\delta(t-t')\tilde R_{ij}(\mathbf k) , \\
F^{\left(lm\right)}(\mathbf k)&=\sum_{i,j=1}^{4}\eta_i^{\left(l\right)}(\mathbf k)\eta^{\left(m\right)}_j(-\mathbf k)\tilde R_{ij}(\mathbf k) .
\label{Fij}
\end{align}
The matrix elements $\tilde R_{ij}(\mathbf k)$ and therefore $F^{\left(ij\right)}(\mathbf k)$ for the MPC fluid can be calculated using Eqs.~(\ref{FDT}) and (\ref{eta_ijkl}).
The expressions for $F^{\left(ij\right)}(\mathbf k)$, and thereby the relevant correlation functions $C_{ij}(\mbf k,\dot\gamma)$, are identical
to those presented in Ref.~\cite{lutsko1985}, with the isentropic coefficients replaced by isothermal counterparts, even though the stress tensor of the  MPC fluid is asymmetric.
With the explicit form of $F^{\left(ij\right)}(\mathbf k)$, $\boldsymbol \eta(\mathbf k)$, and $\boldsymbol \xi(\mathbf k,t)$, the density and velocity correlations and the coupling
between the transverse velocity components can be written as~\cite{lutsko1985}
\begin{equation}
C_{ii}(\mathbf k,\dot\gamma)=C_{ii}(\mathbf k,0)\left[1+\Delta_{ii}\left(\mathbf k,\dot\gamma\right)\right]~,
\label{c_ij_fin}
\end{equation}
and
\begin{equation}
C_{34}(\mathbf k,\dot\gamma)=\frac{k_BT}{\rho_0}\Delta_{34}\left(\mathbf k,\dot\gamma\right)~,
\label{c_34}
\end{equation}
where the equilibrium correlations are
$C_{11}(\mathbf k,0)=\rho_0 k_BT/c_T^2$, with $c_T$  the isothermal speed of sound, and $C_{ii}(\mathbf k,0)=k_BT/\rho_0$ for $i=2,3,4$, and $C_{ij}(\mathbf k,0)=0$ for $i\neq j$~\cite{varghese2015,huang2012}.
The nonequilibrium contributions $\Delta_{ij}(\mathbf k,\dot\gamma)$ are given by
\begin{align}
\Delta_{11}&=\Delta_{22}=-\dot\gamma\int_{0}^{\infty}dt~\frac{kk_x k_y(-t)}{k^3(-t)}e^{-\tilde\nu\chi(\mathbf k,t)}\label{c22} ,\\
\Delta_{33}&=2\dot\gamma\int_{0}^{\infty}dt~\frac{k_xk_y(-t)}{k^2}e^{-2\nu\chi(\mathbf k,t)}\label{c33} ,\\
\Delta_{34}&=\dot\gamma\int_{0}^{\infty}dt~\left[\frac{2k_xk_y(-t)}{kk(-t)}F(\mathbf k,t)-\frac{k_z}{k}\right]e^{-2\nu\chi(\mathbf k,t)} , \label{c34}
\end{align}
with the kinematic viscosity $\nu=\eta/\rho_0$, the sound attenuation factor $\tilde\nu=(\eta+\eta^k/3)/\rho_0$, and
\begin{align}
\chi(\mathbf k,t)&=k^2t+\dot\gamma k_xk_yt^2+\frac{1}{3}\dot\gamma^2k_x^2t^3~,\\
F(\mathbf k,t)&=M\left(\mathbf k\left(-t\right)\right)-\frac{k\left(-t\right)}{k}M(\mathbf k)~ .
\end{align}
Here,
\begin{align}
M(\mathbf k)&=-\frac{kk_z}{k_xk_\perp}\arctan\left(\frac{k_y}{k_\perp}\right)\label{M_def}
\end{align}
and $k_\perp^2=k_x^2+k_z^2$. We do not provide the expression for $\Delta_{44}$, which corresponds to the transverse velocity correlations in the
direction perpendicular to the gradient direction, as the corresponding simulation data suffer from large statistical errors.
We note that for an incompressible fluid $\Delta_{11}$ and $\Delta_{22}$ vanish, while $\Delta_{33}$ retains the form as in Eq.~(\ref{c33})~\cite{wada2004,zarate2009}.
The simulation results for the transverse velocity component, as will be discussed in the following section, will therefore also hold for incompressible fluids.
\begin{figure}[t]
\includegraphics[width=\columnwidth]{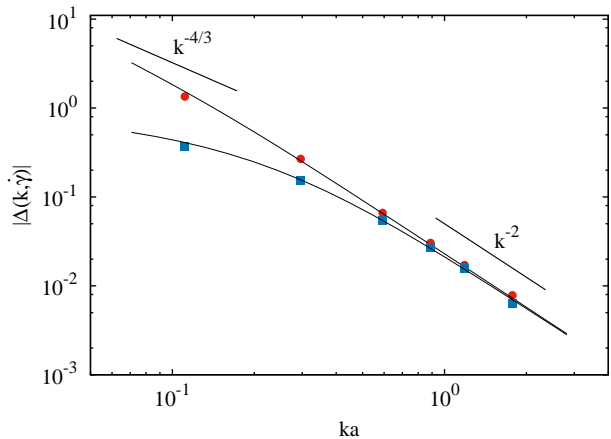}
\caption{The nonequilibrium contribution to the transverse ($\Delta_{33}$, top) and longitudinal ($\Delta_{22}$, bottom) velocity correlations for $k_x=k_y$ and $k_z=0$. The lines correspond to
the theoretical predictions given by Eqs.~(\ref{c22}) and (\ref{c33}), and the symbols indicate simulation results. Parameters: $\dot\gamma=0.04\tau^{-1}$ and $h=0.1\tau$.
}
\label{fig1}
\end{figure}
\begin{figure}[t]
\includegraphics[width=\columnwidth]{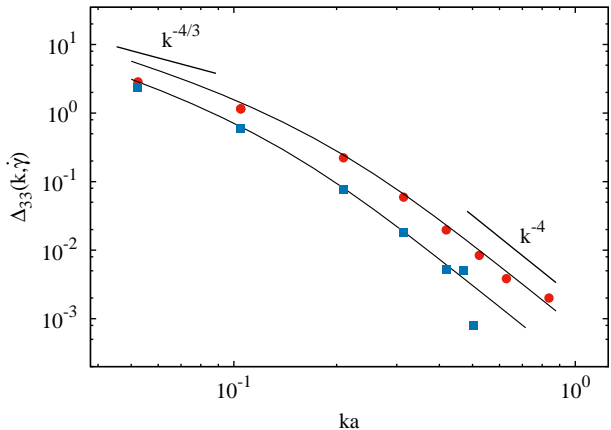}
\caption{The nonequilibrium contribution to the transverse velocity ($\mathbf e^{(2)}$ direction) correlations for $k_y=k_z=0$.
The line corresponds to the theoretical prediction as given in Eq.~(\ref{c33}) and points represent the simulation data.
Parameters: $h=0.2\tau$ and $\dot\gamma=0.02\tau^{-1}$ (top) and $\dot\gamma=0.01\tau^{-1}$ (bottom)}.
\label{fig2}
\end{figure}

\section{Simulation approach}

\subsection{MPC simulations}{\label{simulations}}
In MPC simulations, the fluid is represented by point particles and their positions and velocities evolve in alternating discrete steps -- streaming and collision.
In the streaming, the particles moves ballistically, i.e., the positions are updated as
\begin{equation}
\mathbf r_i(t+h)=\mathbf r_i(t)+h\mathbf v_i(t) ,
\end{equation}
where $h$ is the time step.
In the collision, the particles are grouped into cubic cells of length $a$, and a stochastic rotation of the relative velocities, with respect to the center-of-mass velocity, of the particles is performed in each cell, i.e.,
\begin{equation}
\mathbf v_i(t+h) =\bm{v}_{cm}(t)+\mathcal R\left(\alpha\right)\left(\mathbf v_i(t)-\bm v_{cm}(t)\right)~,
\end{equation}
where
\begin{equation}
\bm{v}_{cm}=\frac{1}{N_c}\sum_{i\in \text{cell}}\mathbf v_i~.
\end{equation}
Here, $N_c$ is the number of particles in the cell which contains particle $i$.
$\mathcal R\left(\alpha\right)$ is the rotation matrix around a randomly oriented axis and $\alpha$ is the rotation angle \cite{kapral2008,gompper2009}.
The shear flow is implemented by Lees-Edwards boundary conditions, in which the flow is generated by moving periodic simulation boxes by a velocity proportional to the boxes' vertical position compared to the primary box~\cite{lees1972,allen1989}.  This boundary conditions is well suited for studying bulk properties of sheared fluids, as it eliminates finite size effects due to solid boundary walls.
In order to maintain an isothermal state under shear, we employ a cell-level Maxwell-Boltzmann rescaling
of the relative velocities of the particles~\cite{huang2010}.
This rescaling method for MPC fluid has been shown to reproduce the isothermal states consistent with the fluctuating Navier-Stokes equations in equilibrium and under shear flow~\cite{huang2012,varghese2015}.

Length- and time-scales in our simulations are given in terms of the length $a$ of a cubic cell and $\tau=\sqrt{ma^2/k_BT}$, where $m$ is the mass of a MPC fluid particle.
The rotation angle $\alpha$ is chosen as $130^\circ$, the average number of fluid particles per cell as $\langle N_c \rangle =10$,  and the time steps $h=0.1\tau$ and $0.2\tau$ are used.
The length of the simulation box ranges from $L=20a$ to $120a$.
The values of the transport coefficients for a given set of simulation parameters
can be evaluated from theoretical expressions~\cite{tuezel2006,gompper2009,winkler2009} and are given by $\nu=0.870a^2/\tau$ and $\tilde\nu=0.886a^2/\tau$ for $h=0.1\tau$, and $\nu=0.508a^2/\tau$
and $\tilde\nu=0.540a^2/\tau$ for $h=0.2\tau$.

\subsection{Simulation results and discussion}

In order to calculate the correlations in the simulations, the velocity field of the fluid in Fourier space is defined as
\begin{equation}
\delta \tilde{\mathbf u}(\mathbf k)=\frac{m}{\rho_0}\sum_{i=1}^{N}[\mathbf v_i-\dot\gamma y_i \hat{\mathbf x}]e^{-i\mathbf k\cdot \mathbf r_i} ,
\label{uk_def_sim}
\end{equation}
where $\rho_0=mN/V$ is the mean mass density and $V=L^3$. The mean flow is subtracted from the particle velocity, hence $\delta \tilde{\mathbf u}$ represents the thermally  fluctuating part.
The velocity correlation functions are then evaluated as
\begin{equation}
C_{ij}(\mathbf k, \dot\gamma)=V^{-1}\langle \delta \tilde u^{(i-1)}(\mathbf k)\delta \tilde u^{(j-1)}(-\mathbf k)\rangle~,
\label{corr_def_sim}
\end{equation}
where $i,j\in\{2,3,4\}$. $C_{22}$ corresponds to the longitudinal, and $C_{33}$ and $C_{44}$ to the transverse components.
At equilibrium, i.e., $\dot\gamma=0$, the velocity correlation functions with velocity fields defined as in Eq.~(\ref{uk_def_sim}) are given by
$C_{ii}=k_BT/\rho_0$ and $C_{ij}=0$ for $i\neq j$.
In order to avoid $\mathcal O(N^2)$ computational time in evaluating correlation functions in shear flow implemented by Lees-Edwards boundary conditions~\cite{lange2009},
the sampling is carried out at every $1/\dot\gamma$ time steps, at which the usual periodic boundary conditions apply ($\mathcal O(N)$ computational time).
\subsubsection{Velocity fluctuations}{\label{simulation_velocity}}
\begin{figure}[t]
\includegraphics[width=\columnwidth]{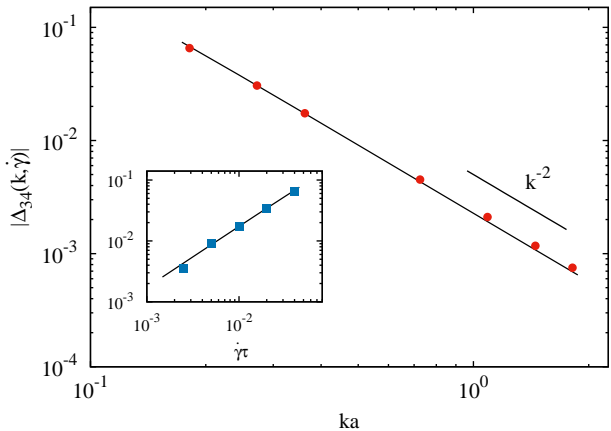}
\caption{The coupling between the transverse velocity components for $k_x=k_y=k_z>0$. The inset shows the variation of the coupling with the shear rate for $k_z=2\pi/60$.
The lines represent the theoretical predictions by Eq.~(\ref{c34}) and the simulation data are represented by points.
Parameters: $h=0.1\tau$ and $\dot\gamma=0.04\tau^{-1}$ for the main  plot.}
\label{fig3}
\end{figure}
As is well known, the velocity fluctuations in a fluid at equilibrium are isotropic and spatially delta-correlated.
Thus, in the Fourier-space representation, the correlations are independent of the wave vector $\mathbf k$.
However, as is evident from Eqs.~(\ref{c22}) and (\ref{c33}), the nonequilibrium contribution $\Delta_{ij}$ to the correlations 
in shear flow depends on the magnitude as well as the direction of the wave vector, thereby rendering the fluctuations spatially correlated and anisotropic.
Figure \ref{fig1} displays the nonequilibrium contributions
of the longitudinal ($\mathbf e^{(1)}$ direction) and transverse ($\mathbf e^{(2)}$ direction)
velocity correlations for a wave vector in the flow-gradient plane pointing at an angle $45^\circ$ to the flow direction.
The correlations decay as $k^{-2}$ for large wave vectors.
This power-law dependence for the transverse velocity correlations has been derived previously~\cite{zarate2008}.

Figure 2 displays the nonequilibrium enhancement of transverse velocity correlations for the wave vector pointing along the flow direction, i.e., $k_y=k_z=0$.
As predicted by the theory, the correlations decay as $k^{-4}$ for large wave vectors.
Since the $k^{-4}$ dependence corresponds to diverging correlations in real space, a slower decay is expected for smaller wave vectors.
From the asymptotic analysis of the correlation functions,
it was shown that the velocity correlations decay as $k^{-4/3}$ for small $k$~\cite{wada2003,wada2004,zarate2008}.
This power-law dependence corresponds to a $r^{-5/3}$ decay in real space and is independent of the direction of the wave vector~\cite{zarate2008}.
Unfortunately, we are not able to probe this small wave vector decay due to limitations in system size and shear rate in our simulations.
However, it is clear from our simulation results that the velocity correlations indeed show a cross over from  $k^{-4}$ (Fig.~\ref{fig2}) and also $k^{-2}$ (Fig.~\ref{fig1}) decay to a slower decay for small wave vectors.
The agreement between the simulation results and the theoretical expressions for the  wave vectors accessible for finite-system sizes strongly suggest that the correlations for infinite systems would indeed be long-ranged.
It is noteworthy that rigid boundaries, which are absent in our simulations, is
expected to modify the $k^{-4/3}$ small wave vector decay of the velocity correlations to a $k^2$ dependence~\cite{zarate2008,zarate2009,zarate2006} .

The transverse components of the fluid-velocity field  at equilibrium are well known to be identical and decoupled.
Under shear, the degeneracy is lifted and the components become coupled~\cite{lutsko1985,varghese2015}. The strength of the coupling is given by Eq.~(\ref{c34}).
Figure~\ref{fig3} shows the variation of the coupling for the wave vector pointing out of the shear-gradient plane.
The coupling decays as $k^{-2}$, corresponding to $r^{-1}$ decay, for the entire range of wave vectors
in our simulations. In agreement with the theory,
the coupling increases linearly with shear rate (cf. inset of Fig.~\ref{fig3}). As is evident from Eq.~(\ref{M_def}), the coupling vanishes for wave vectors in the shear-gradient plane ($k_z=0$).

For large shear rates and small wave vectors, we observe deviations for the velocity correlations obtained by simulations and predicted theoretically (Fig.~\ref{fig2} top curve).
We notice that a similar unexpected behavior was found for the large-length-scale decay of real-space correlations in sheared granular fluids~\cite{otsuki2009-pre}.
This effect may be attributed to a density dependence of the viscosity of the fluid.
For fluids with density-dependent viscosity, the density fluctuations will be coupled to the mean flow velocity through an additional
term in the linearized Navier-Stokes equation.
This coupling is non-negligible for high shear rates and
modify the eigenvalues and eigenvectors and thereby the hydrodynamic fluctuations.
We hope to address this issue in detail in the future.

\subsubsection{Density correlations}{\label{simulation_density}}
\begin{figure}[t]
\includegraphics[width=\columnwidth]{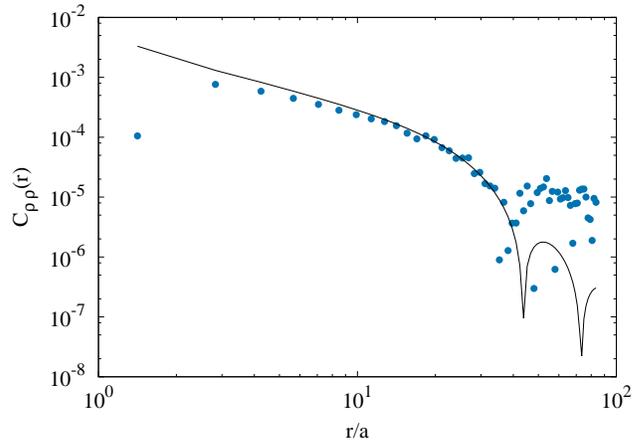}
\caption{Spatial correlations in density fluctuations in shear flow. The line represents Eq.~(\ref{corr_rho_sum}) and symbols simulation results.
The correlations are measured in the shear-gradient plane and in the direction $45^\circ$ to the flow.
Parameters: $h=0.2\tau$, $\dot\gamma=0.01\tau^{-1}$ and $L=120a$.}
\label{fig4}
\end{figure}
The density fluctuations, similar to the velocity fluctuations, become spatially correlated in shear flow. The
correlation in the density fluctuations is defined as
\begin{equation}
C_{\rho\rho}(\mathbf r)=\langle \delta \rho(\mathbf r)\delta\rho(\mathbf 0)\rangle~,
\label{corr_rho_def}
\end{equation}
where $\delta\rho(\mathbf r)=\rho(\mathbf r)-\rho_0$. In simulations, the correlation is measured as $m^2\langle \delta n(\mathbf r_0)\delta n(\mathbf r_0+\mathbf r)\rangle$,
where $n(\mathbf r)$ is number of MPC particle in a given cell, with $\mathbf r_0$ and $\mathbf r$ taken as the centers of the cells.
The theoretical expressions for $C_{\rho\rho}(\mathbf r)$ can be obtained by inverting $C_{11}(\mathbf k,\dot\gamma)$ (see Eq.~(\ref{c_ij_fin}) and subsequent discussion), and  is given by
\begin{equation}
C_{\rho\rho}(\mathbf r)=\frac{m \rho_0}{V}\sum_{\mathbf k}\left[1+\Delta_{11}(\mathbf k,\dot\gamma)\right]\cos(\mathbf k\cdot\mathbf r)~,
\label{corr_rho_sum}
\end{equation}
where $\Delta_{11}(\mathbf k,\dot\gamma)$ is given by Eq.~(\ref{c22}).

Figure~\ref{fig4} shows a comparison of the MPC simulation results for the density correlations and the numerically
evaluated theoretical expression ~(\ref{corr_rho_sum}).
The agreement between the two results is good for moderate length scales.
The deviation of the simulation results for small length scales is firstly due to the fact that the theoretical expression for the nonequilibrium contribution $\Delta_{11}(\mathbf k,\dot\gamma)$ is
a good approximation for small wave vectors only. Secondly, the validity of Navier-Stokes equations for a MPC fluid breaks down at small length scales $r\lesssim \pi\sqrt{\nu h}$ \cite{huang2012}.
One the other hand, for large length scales, the simulation data suffer from large statistical errors.
However, for intermediate length scales, the correlations are well reproduced.

\subsubsection{Pressure fluctuations}{\label{simulation_pressure}}
\begin{figure}[t]
\includegraphics[width=\columnwidth]{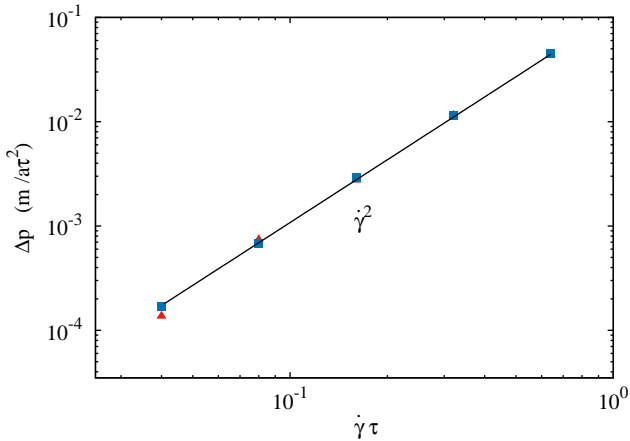}
\caption{The nonequilibrium contribution to the pressure tensor (Eq.~(\ref{pressure})), for the gradient (triangles) and vorticity (squares) components.
Parameters: $L=60a$, $h=0.1\tau$.}
\label{fig5}
\end{figure}
The nonequilibrium contribution to the velocity correlations is also manifested in a shear-rate dependence of the pressure of the fluid~\cite{wada2003}.
The pressure in our simulations was measured as time average of the diagonal components of the instantaneous pressure tensor~\cite{winkler2009}
\begin{equation}
p_{\alpha\beta}=\frac{m}{V}\sum_{i=1}^{N} \left[v_{i\alpha}v_{i\beta}+\frac{1}{h}\Delta v_{i\alpha} r_{i\beta}'+\delta_{\alpha x}\delta_{\beta y}\frac{\dot\gamma h}{2}v_{iy}^2\right]
\label{pressure}
\end{equation}
where ${\bm v}_i=\mathbf v_i-\dot\gamma y_i \hat{\mathbf x}$ is the thermal velocity after streaming,
$\Delta \bm v_i$ is the change in the velocity in the collision step, and $\mathbf r_i'$ is the position of the particle in the
grid-shifted frame in the collision step.
The pressure tensor for a  MPC fluid at equilibrium is given by the ideal gas equation of state $p_{\alpha\beta}=\delta_{\alpha\beta}n_0 k_BT$, where $n_0$ is the mean number density.

Figure~\ref{fig5} displays the deviation of the diagonal components of the pressure tensor from the equilibrium value.
We observe that for high shear rates, the deviations vary as $\dot\gamma^\zeta$, with $\zeta=2$.
The exponent $\zeta$ associated with the change in pressure has been controversial. Mode coupling theory~\cite{kawasaki1973}
predicts $\zeta=3/2$, which has been confirmed in simulations of Lenard-Jones fluids at the triple point \cite{evans1981}.
However, in atom-scale simulations using two- and three-body potentials, the exponent was observed to be $\zeta \approx 2.0$ away from the triple point~\cite{marcelli2001pre}. 
The deviation was ascribed to two-body interactions~\cite{marcelli2001fluid,ge2001}. 
Further systematic studies using the Lenard-Jones potential found the exponent in the range $\zeta =1.2  - 2.0$ as a function of density and temperature~\cite{ge2003},
thereby rendering $\zeta=3/2$ only a special case.
Fluctuating hydrodynamic calculations of incompressible fluids on the other hand predict two limiting regimes of the variation of pressure,
depending on the value of the dimensionless parameter
$\lambda=\dot \gamma L^2/\nu$~\cite{wada2003}. The exponent is expected to be $\zeta=2$ for $\lambda\ll 1$ and $\zeta=3/2$ for $\lambda\gg 1$.
However, our simulations yield $\zeta=2$ even for $\lambda\gg1$, which is not a contradiction, as MPC fluid is highly compressible.
It therefore remains for further theoretical and simulation studies to establish a unified picture of the 
exponent associated with the hydrostatic pressure under shear.

\section{Conclusions}{\label{conclusions}}
Using MPC simulations, we have unambiguously demonstrated that the hydrodynamic fluctuations in simple fluids under shear flow, in contrast to the fluctuations at equilibrium, are spatially correlated.
The correlations in velocity and density fluctuations are shown to be anisotropic and spatially correlated over the entire volume of the system.
For large wave vectors, the decay of the velocity correlations exhibits a power-law dependence in the wave vector.
We observe a $k^{-4}$ decay for  wave vector pointing along the flow direction and a $k^{-2}$ dominated decay along off-flow directions.
For small wave vectors, we observe a cross-over to slower decay, indicating algebraic decay of real-space correlations at large distances.
In contrast to the case at equilibrium, the transverse velocity components are coupled under shear, with a $k^{-2}$ dependence for the entire range of wave vectors accessible in our system,
corresponding to $r^{-1}$  real space correlations.
We find good agreement between the simulation results and the predictions of fluctuating hydrodynamics without any fitting parameters.

Although we have shown that the hydrodynamic fluctuations under shear are spatially correlated, it remains for further simulation studies to show they are truly long-ranged, i.e., for instance,
that the transverse velocity fluctuations decay as $k^{-4/3}$ for small wave vectors.
This demands either higher shear-rates or large system sizes than those accessible in our current study.
Since MPC fluid is compressible with a density-dependent viscosity, large density fluctuations -- although they are physical -- arise
under such conditions. Other hydrodynamic simulation methods for incompressible fluids, which incorporate thermal fluctuations, may therefore alternatively prove adequate.

We also find in our simulations that the nonequilibrium contribution to the hydrostatic pressure
follows a $\dot\gamma^2$ dependency on shear rate, in contrast to the $\dot\gamma^{3/2}$ dependency predicted by mode-coupling theory.
Our mesoscale simulation result is in agreement with the findings of previous atom-scale simulations studies using a range of interaction potentials, 
where deviation from the $\dot \gamma^{3/2}$ dependency was observed. 
A fundamental understanding of the hydrostatic pressure under shear is still lacking.
The effect of confining walls on pressure and spatial correlations in hydrodynamic fluctuations are also yet to be addressed in simulations.

%

\end{document}